\documentclass[prd, nobibnotes, nofootinbib, eqsecnum, twocolumn,
10pt, a4paper, superscriptaddress, nofootinbib, showpacs,
byrevtex]{revtex4}

\usepackage{psfig}

\newcommand{\ETAL}{{\it et al.}}

\newcommand{\Hconf}{{\cal H}}
\newcommand{\BAR}{{\rm b}}
\newcommand{\MAT}{{\rm mat}}
\newcommand{\SCAL}{{\rm S}}
\newcommand{\DM}{{\rm \scriptscriptstyle DM}}
\newcommand{\SDMDM}{ { {\rm \scriptscriptstyle DM}
                       -
                       {\rm \scriptscriptstyle DM} } }

\newcommand{\SGAMMADM}{ { \gamma
                          -
                          {\rm \scriptscriptstyle DM} } }
\newcommand{\SNUDM}{ { \nu
                     -
                     {\rm \scriptscriptstyle DM} } }
\newcommand{\DMGAMMA}{${\rm DM} - \gamma$}
\newcommand{\GAMMADM}{$\gamma - {\rm DM}$}
\newcommand{\GAMMAE}{$\gamma - e$}
\newcommand{\IDM}{{\rm \scriptscriptstyle IDM}}
\newcommand{\CDM}{{\rm \scriptscriptstyle CDM}}
\newcommand{\WDM}{{\rm \scriptscriptstyle WDM}}
\newcommand{\THM}{{\rm Th}}
\newcommand{\STRUCT}{{\rm struct}}

\newcommand{\DEC}{{\rm dec}} 
 
\newcommand{\SOUND}{{\rm s}} 

\newcommand{\UUNIT}[2]{\,{\rm #1}^{#2}}
\newcommand{\UNIT}[2]{{\rm #1}^{#2}}
\newcommand{\AVRG}[1]{\left<{#1}\right>}
\newcommand{\EE}[2]{{#1} \times 10^{#2}}
\newcommand{\UEE}[1]{10^{#1}}

\newcommand{\DD}{w}

\begin{document}

\title{Interacting Dark Matter disguised as Warm Dark Matter}

\author{C\'eline B\oe{}hm}

\email{boehm@astro.ox.ac.uk}

\affiliation{Department of Physics, Nuclear \& Astrophysics Laboratory,
  University of Oxford, Keble Road, Oxford OX1 3RH, United Kingdom}
\affiliation{L.P.M.T., 5 place Eug\`ene Bataillon, F--34095 Montpellier
  II, France}

\author{Alain Riazuelo}

\email{riazuelo@spht.saclay.cea.fr}

\affiliation{Service de Physique Th\'eorique, CEA/DSM/SPhT, \\
Unit\'e de Recherche associ\'ee au CNRS, CEA/Saclay, F--91191
Gif-sur-Yvette c\'edex, France}
\affiliation{D\'epartement de Physique Th\'eorique, Universit\'e de
  Gen\`eve, 24, Quai Ernest Ansermet, CH--1211 Gen\`eve 4,
  Switzerland}

\author{Steen H.~Hansen}

\email{hansen@astro.ox.ac.uk}

\affiliation{Department of Physics, Nuclear \& Astrophysics Laboratory,
  University of Oxford, Keble Road, Oxford OX1 3RH, U.K.}

\author{Richard Schaeffer}

\email{rschaeffer@cea.fr}

\affiliation{Service de Physique Th\'eorique, CEA/DSM/SPhT, \\
Unit\'e de Recherche associ\'ee au CNRS, CEA/Saclay, F--91191
Gif-sur-Yvette c\'edex, France}

\date{11 September 2002}

\begin{abstract}
We explore some of the consequences of Dark-Matter--photon
interactions on structure formation, focusing on the evolution of
cosmological perturbations and performing both an analytical and a
numerical study.  We compute the cosmic microwave background
anisotropies and matter power spectrum in this class of models.  We
find, as the main result, that when Dark Matter and photons are
coupled, Dark Matter perturbations can experience a new damping regime
in addition to the usual collisional Silk damping effect. Such Dark
Matter particles (having quite large photon interactions) behave like
Cold Dark Matter or Warm Dark Matter as far as the cosmic microwave
background anisotropies or matter power spectrum are concerned,
respectively. These Dark-Matter--photon interactions leave specific
imprints at sufficiently small scales on both of these two spectra,
which may allow us to put new constraints on the acceptable
photon--Dark-Matter interactions.  Under the conservative assumption
that the abundance of $10^{12} M_\odot$ galaxies is correctly given by
the Cold Dark Matter, and without any knowledge of the abundance of
smaller objects, we obtain the limit on the ratio of the
Dark-Matter--photon cross section to the Dark Matter mass
$\frac{\sigma_\SGAMMADM}{m_\DM} \lesssim 10^{-6}
\frac{\sigma_\THM}{100 \UUNIT{GeV}{}} \simeq \EE{6}{-33} \UUNIT{cm}{2}
\UUNIT{GeV}{-1}$.

\vspace*{0.5cm}
\noindent
{\footnotesize Preprint numbers: SPhT-Saclay T01/147, {\tt
astro-ph/0112522}}
\end{abstract}


\maketitle

\section{Introduction}

The nature of Dark Matter particles remains one of the major
challenges for both fundamental physics and astrophysics. Whereas Cold
Dark Matter (CDM) perfectly explains the formation of large scale
structure~\cite{Peacock:1999ye} on scales greater than $1
\UUNIT{Mpc}{}$, there seem to be various discrepancies on smaller
(subgalactic) scales. Some of these come from the following.
\begin{enumerate}  
  
\item \label{prob1} $N$-body CDM simulations, which give cuspy halos
  with divergent profiles toward the center~\cite{NFW}, in potential
  disagreement with the galaxy rotation curves~\cite{binney} and with
  observations from gravitational lensing~\cite{flores};
  
\item \label{prob2} Bar stability in high surface brightness spiral
  galaxies which also demands low-density cores~\cite{debattista};
  
\item \label{prob3} CDM models which for years have been seen to
  yield an excess of small scale structures~\cite{ss1985}.  Numerical
  simulations~\cite{Klypin:1999uc} found $1$--$2$ orders of magnitude
  more satellite galaxies than what is
  observed~\cite{Moore:1999wf}. However, recent
  work~\cite{stoehr,haya,chiu} indicate that there may be no problem
  for the galaxy mass function after all.

\item \label{prob4} The formation of disk galaxy angular momentum,
  which is much too small in galaxy simulations~\cite{NFW}.

\end{enumerate}
Problems~\ref{prob3} and~\ref{prob4} can be solved with the usual Warm
Dark Matter (WDM) which experiences free streaming and hence
suppresses power on small scales~\cite{ss1988,sommerlarsen,nbody}.
Such a particle physics candidate is easy to find in a minimalistic
extension of the Standard Model, namely, a sterile
neutrino~\cite{dodelson,Shi:1999km,Dolgov:2000ew,Abazajian:2001nj}.
However, collisionless WDM does not solve problems~\ref{prob1}
and~\ref{prob2} (see Ref.~\cite{knebe}) and one is therefore forced to
propose more complicated models like scenarios of nonthermal
production of Weakly Interacting Massive Particles
(WIMPs)~\cite{zhang} for instance.

On the other hand, Strongly Interacting Dark Matter (SIDM) has been
suggested~\cite{spergel} to solve problems~\ref{prob1} and~\ref{prob2}
and does so successfully provided the cross section is within the
range $\EE{2}{-25} \UUNIT{cm}{2} \UUNIT{GeV}{-1} \lesssim
\sigma_\SDMDM / m_{\DM} \lesssim \UEE{-23} \UUNIT{cm}{2}
\UUNIT{GeV}{-1}$~\cite{daveS,wandelt} (where $m_\DM$ and
$\sigma_\SDMDM$ are the Dark Matter mass and self-interaction
cross section respectively).  Problem~\ref{prob3} is also partly
solved in this scenario~\cite{daveS} but the survival of galactic
halos exclude the range $\EE{6}{-25} \UUNIT{cm}{2} \UUNIT{GeV}{-1}
\lesssim \sigma_\SDMDM / m_{\DM} \lesssim \EE{2}{-20} \UUNIT{cm}{2}
\UUNIT{GeV}{-1}$~\cite{GnedinO}.  Furthermore, since the inner regions
of massive clusters are elliptical, one must have $\sigma_\SDMDM /
m_\DM \lesssim \EE{3}{-26} \UUNIT{cm}{2}
\UUNIT{GeV}{-1}$~\cite{miralda}.  One therefore concludes that the
allowed cross section is slightly too small and that SIDM, on its own,
cannot solve problems~\ref{prob1}--\ref{prob4}.

This paper is finally motivated by the recent findings~\cite{bfs2001}
that either Dark-Matter--photon or Dark-Matter--neutrino interactions
can transfer to Dark Matter the damping that the photon or neutrino
fluids undergo. This process, characterized in the simplest cases by
an exponential cutoff in the matter power spectrum, was referred to as
``induced damping'' in Ref.~\cite{bfs2001}. In particular, by
requiring that the damping induced by relativistic particles does not
wash out the Dark Matter primordial fluctuations responsible for the
formation of the smallest galaxies, it was found that the ratio of the
corresponding cross sections to the Dark Matter mass must satisfy
$\sigma_\SGAMMADM / m_\DM <
\UEE{-30} \UUNIT{cm}{2} \UUNIT{GeV}{-1}$ and $\sigma_\SNUDM / m_\DM <
\UEE{-34} \UUNIT{cm}{2} \UUNIT{GeV}{-1}$.  It was then suggested that,
at the edge to be satisfied, these constraints could provide an
alternative scenario for Warm Dark Matter.

However, the exact value of these cross sections as well as the shape
of the resulting power spectrum depend on the details of the
interactions history of the fluids. We therefore determine---in this
paper---the transfer function resulting from non-negligible
interactions between Dark Matter and photons. The case of
neutrino--Dark-Matter interactions will be examined in a subsequent
paper. The effects thereof are naturally different from those due to
self-interactions. However, one should keep in mind that realistic a
Dark Matter particle probably has interactions both with itself and
with other particles.

In Sec.~\ref{celine_part}, we discuss the motivations for such Dark
Matter particles. In Sec.~\ref{alain_part} we describe the effect of
Interacting Dark Matter\footnote{In the following, we shall adopt the
notation of Interacting Dark Matter for convenience but the reader has
to remember that only Dark-Matter--photon interactions are
investigated in this paper.} (IDM) on the early evolution of
cosmological perturbations and, in Sec.~\ref{steen_part}, we give an
analytical fit to the main features produced by IDM on the matter
power spectrum. In the Conclusion, we discuss the main result of our
work, namely that, for Dark Matter coupled to photons, the original
fluctuations are damped as soon as they enter the horizon, because of
the impeded growth due to this coupling. We also discuss a few
implications of this result.

\section{Motivations}
\label{celine_part}

Until the recent Dark Matter crisis, it was very well known that
weakly interacting particles with a mass greater than a few
$\UNIT{keV}{}$ (such as supersymmetric
particles~\cite{blument1,blument2,blument3,blument4} for instance) did
not suffer from prohibitive free streaming~\cite{davis} or collisional
damping effects~\cite{silk68} and could therefore represent a very
promising solution to the Dark Matter puzzle. In fact, the
interactions of such particles are generally assumed to be so weak
that they can be neglected as far as structure formation is concerned.

However, the precise order of magnitude of the Dark Matter
interactions for which it is justified to neglect these damping
effects has never been given explicitly.  A hint at the answer comes
from investigating both the free streaming and collisional damping
scales of Dark Matter primordial fluctuations, taking into account all
the possible interactions. By requiring that the latter do not wash
out the fluctuations responsible for the formation of the smallest
primordial structures [$M_\STRUCT \sim (10^6$--$10^9) M_\odot$], one
can obtain bounds on the Dark Matter particle's mass and interaction
rates.

Let us consider primordial fluctuations made of Dark Matter and
ordinary species like photons, neutrinos, baryons, and electrons,
etc. One can show that the largest collisional damping effects may be
due to the Dark Matter interactions with relativistic particles (e.g.,
neutrinos and photons).  Focusing on the Dark-Matter--photon
interactions, an analytic calculation~\cite{bfs2001} shows that Dark
Matter must decouple from the photons at a redshift $z_\DEC$ greater
than $\sim 10^5$ to ensure that the spectrum at scales $k \le
k_\STRUCT \sim 10$--$100 \UUNIT{Mpc}{-1}$ (corresponding to the mass
$M_\STRUCT$ given above) is not exponentially damped by the photon
interactions with baryons and electrons.  Because of the exponential
nature of the collisional damping effect, this necessary condition
holds whatever the amplitude of the initial fluctuation spectrum is,
and whatever its past history is.

The bound on $z_\DEC$ can actually be translated into a limit on the
Dark-Matter--photon interaction rate which finally turns into a
constraint on the ratio of the photon--Dark-Matter cross section (at
the \DMGAMMA{} decoupling) to the Dark Matter mass:
$\AVRG{\sigma_\SGAMMADM v } / (m_\DM c) < \UEE{-30} \UUNIT{cm}{2}
\UUNIT{GeV}{-1}$ (here the angular brackets denote the statistical
average owing to the fact that the coupling is due to momentum
transfer).  Thus, the lower the cross sections are, the smaller the
Dark Matter mass must be in order to maintain the thermal equilibrium
a for very long time.

The properties of such interacting Dark Matter have already been
discussed in Ref.~\cite{bfs2001} (and will be so in more details
in~\cite{bfs2002}).  In particular, it was pointed out that if they
thermally decouple at a redshift\footnote{The value of the redshift
corresponding to the Dark-Matter--photon decoupling will be determined
properly in this paper.} close to $z \sim 10^5$ they may behave as WDM
particles erasing structures with a size smaller than $10$--$100
\UUNIT{kpc}{}$.  Many other cases of WDM were considered in these
papers but that of large Dark-Matter--photon interactions is
especially interesting because one may expect some modifications in
both the Cosmic Microwave Background (CMB) and matter power spectra.
 
Despite their rather strong interactions with photons, these Dark
Matter particles may still be considered as ``dark'' particles. Their
thermal decoupling indeed occurs much before the recombination epoch
and they therefore keep the Universe transparent to photons from the
last scattering surface to nowadays (as in the standard scheme).
However, one may wonder if this Dark Matter is able to accumulate into
stars and whether or not it may affect their properties.  Note that,
for the cross sections mentioned above, Dark Matter particles---once
thermalized---have a mean free path $\lambda_\DM \sim (m_\DM / 1
\UUNIT{GeV}{})^{1/2} R_\odot$ within the sun, potentially giving rise
to heat conduction if their mass is smaller than a few $\UNIT{GeV}{}$.
For $m_\DM > m_p \sim 1 \UUNIT{GeV}{}$~\cite{gould}, on the other
hand, one expects Dark Matter to be able to evaporate so that no Dark
Matter particles would be left in the sun.  A \GAMMADM{} cross section
$\sim 100$ times lower (as we shall be led to consider below) would
predict even less accumulation in the sun.

A lower limit on the Dark Matter mass can be inferred from the
free streaming constraint.  As far as our specific interaction rates
are concerned, the well known condition $m_\DM > 1
\UUNIT{keV}{}$~\cite{peebles} (obtained for a weakly-interacting
species) has to be replaced~\cite{bfs2001} by $m_\DM > 1
\UUNIT{MeV}{}$.  These limits, however, hold as long as the thermal
decoupling of Dark Matter occurs before the gravitational collapse and
would be disregarded if we were considering at the same time extremely
strong Dark Matter self-interactions for instance (see the discussion
in~\cite{bfs2001} for the exact conditions of validity).\footnote{It
should be stressed that in our specific model the thermal decoupling
of Dark Matter is fixed by the Dark-Matter--photon decoupling.}
  
The condition $m_\DM > 1 \UUNIT{MeV}{}$ also ensures that the number
of relativistic degrees of freedom during the primordial
nucleosynthesis is the usual one.\footnote{Open windows for $m_\DM
\lesssim 1 \UUNIT{MeV}{}$ would require a more involved
scenario~\cite{bfs2002}.  Together with galaxy dynamic results, this
would make such a case very unlikely.}  Hence, one does not expect
any problem concerning primordial nucleosynthesis in this case.

Dark Matter particles must also have an acceptable relic abundance.
This requirement actually constrains the nature of Dark Matter.  We
are considering, indeed, large elastic cross sections from $\UEE{-33}$
to $\UEE{-27} \UUNIT{cm}{2}$ for a Dark Matter mass in the
$\UNIT{MeV}{}$--$\UNIT{TeV}{}$ range for instance (according to the
previous analytical estimate~\cite{bfs2001}). If there exists any
symmetry between elastic and annihilation cross sections, this may
provide annihilation cross sections much larger than the ones required
for Weakly Interacting Massive Particles which are expected to be
roughly of the order of $\UEE{-36} \UUNIT{cm}{2}$ (for particles
having chemically decoupled at their nonrelativistic transition).
However, this constraint is obtained by assuming that the number of
Dark Matter particles is exactly equal that of anti-Dark-Matter
particles in the primordial Universe (or assuming that Dark Matter
particles are Majorana particles). It can be disregarded if one makes
the assumption that there exists a primordial asymmetry between
particles and anti particles so that large elastic
photon--Dark-Matter cross sections may finally be relevant.  Such
Dark Matter particles would then behave like baryons but with smaller
interaction rates and would probably be neutral to avoid important
reionization effects. This should therefore exclude, in principle,
tree-level elastic and annihilation cross sections between fundamental
Dark Matter particles and photons.

On the other hand, it is quite interesting to note that these
constraints on the elastic cross sections from structure formation
potentially imply annihilation cross sections (into two photons) close
to the one proposed in Ref.~\cite{kaplinghat} to make strongly
annihilating Dark Matter a possible solution to the CDM
crisis.\footnote{However, as we will see in the next section, the
constraints we will finally obtain in this paper are smaller by a
factor $\sim 10^{-3}$.}  This, in fact, still represents an open
possibility as discussed in Ref.~\cite{craigdavis}.

As mentioned previously, large elastic scattering cross sections may
imply large annihilation cross sections. If the latter exceed
$10^{-36} \UUNIT{cm}{2}$, the condition for having an acceptable relic
abundance leads to requiring an initial asymmetry between Dark Matter
particles and anti particles. This asymmetry implies that no anti
particles should be left after the Dark Matter annihilation in the
primordial Universe (unless one imposes a very large, quite unnatural
fine-tuning). Therefore, in this case, there should not be any further
annihilation of Dark Matter particles with their anti particles in the
center of galaxies or in galaxy clusters (similarly to baryons
actually). No extra X-ray emission can then be expected from such Dark
Matter particles (still assuming they have a too large annihilation
cross section). This could in fact provide an important signature.  If
for one reason or another Dark Matter is shown to annihilate in the
galactic center or in cluster of galaxies, one should be able to rule
out such strongly interacting particles (for masses greater than a few
$\UNIT{MeV}{}$ as we will see in the following, depending on the scale
of the structures considered) or at least to strongly constrain their
characteristics.  In any case, it would become necessary to
investigate their properties more closely.
  
On the other hand, one could also relax the assumption of a crossing
symmetry between the elastic and the annihilation cross sections.
However, it is hard in this case to predict whether the annihilation
cross section would be much larger than the elastic cross section or
not since no realistic models exhibit such an asymmetry.

At low energy, the elastic scattering cross sections should not induce
any deviations from the black body spectrum.  In any case, particles
annihilating before (or around) $z \sim 10^{10}$ and thermally
decoupling before $z \sim 10^5$ leave enough time for any irregularity
to be erased at the recombination epoch (because of the usual coupling
between photons and baryons). At least, if some distortions exist,
they should be very small.

Finally, a still important question concerns the signature of such
interactions in the CMB and matter power spectra. This is precisely
the aim of the present paper.  We shall adopt the definitions given
in~\cite{bfs2001}, following which:
\begin{itemize}

\item particles having a collisional damping 
  or free streaming length of the order of $l_\STRUCT \propto
  k^{-1}_\STRUCT$ are called Warm Dark Matter; and

\item particles having a collisional damping or free streaming length 
  much lower than $l_\STRUCT \propto k_\STRUCT^{-1}$ are called Cold
  Dark Matter.
\end{itemize}
This actually differs from the current WDM definition since we now
take into account the possibility that Dark Matter may be ``warm'' not
because of its mass but because of its interactions with another
species (referred to as ``induced damping'' effects
in~\cite{bfs2001}).  Since the above criteria are based on the shape
of the matter power spectrum (not on the form of the CMB spectrum),
the signature of such Interacting Dark Matter in the CMB anisotropies
cannot be easily inferred without performing numerical calculations.
We now investigate how such a kind of Dark Matter changes the relevant
equations for determining both the CMB and the matter power spectra.

\section{The physics of Interacting Dark Matter}
\label{alain_part}

In this section, we first recall the main physical effects that arise
when one considers coupled fluids. As a warm-up, we shall recall the
main equations governing the evolution of the photon--baryon plasma
before recombination (Sec.~\ref{subsecstandardthomson}).  We then
write the modified perturbation equations for the cosmological
perturbations including Interacting Dark Matter
(Sec.~\ref{subsecperteqicdm}) and study the damping experienced by
Dark Matter fluctuations
(Secs.~\ref{subsecdmgamma}--\ref{subsecdamp3}). Finally, the most
prominent observational consequences of IDM on Cosmic Microwave
Background (CMB) anisotropies (Sec.~\ref{subsecCMB}) and on matter
power spectrum (Sec.~\ref{subsecPS}) are discussed.

\subsection{Reminder of the influence of Thomson scattering on
cosmological perturbations}
\label{subsecstandardthomson}

The aim of this paper is to study the interactions between Dark Matter
and photons which are {\it a priori} quite similar to Thomson
scattering between photons and baryons.  Therefore, we shall first
recall the standard case of photons coupled to baryons through Thomson
scattering cross sections and collisionless Dark Matter.  Indeed, as
we will see later, some of the effects that affect the baryon fluid
are also present for Dark Matter.

In the following sections, we will introduce the Euler equation for
photons, baryons, and Dark Matter (Sec.~\ref{subsubseceuler}), and
then compute the Dark Matter perturbation evolution
(Sec.~\ref{subsubsecdm}). We describe a situation in which the
coupling between photons and baryons can play a significant role, and
describe the usual damping phenomena which can affect the cosmological
perturbations (Sec.~\ref{subsubsecdamp}).

\subsubsection{Photon and baryon Euler equations}
\label{subsubseceuler}

We consider nonrelativistic baryons coupled to photons through
Thomson scattering. The corresponding Euler equations for these two
fluids are
\begin{eqnarray}
\label{vbarex}
\dot v_\BAR 
 & = &   k \Phi - \Hconf v_\BAR
       - R^{- 1} \dot \kappa (v_\BAR - v_\gamma) , \\
\label{vphotex}
\dot v_\gamma
 & = &   k \Phi
       + \frac{1}{4} k \delta_\gamma - \frac{1}{6} k \pi_\gamma
       - \dot \kappa (v_\gamma - v_\BAR) ,
\end{eqnarray}
where an overdot denotes a derivative with respect to the conformal
time $\eta$, $\Hconf$ is the conformal Hubble parameter ($\Hconf
\equiv \dot a / a$ with $a$ being the scale factor), $\delta_X$,
$v_X$, and $\pi_X$ represent the density contrast, the velocity
divergence, and the anisotropic stress of the species $X$,
respectively (we work in Newtonian gauge), and $\Phi$ is the Bardeen
potential (see, e.g., Refs.~\cite{pertgen1,pertgen2,pertgen3}). The
Thomson scattering term between photons and baryons reads
\begin{equation}
\label{kappa}
\dot \kappa = a \sigma_\THM n_e ,
\end{equation}
where $\sigma_\THM$ is the standard Thomson scattering cross section,
and $n_e$ is the free electron number density. This quantity is also
referred to as the differential opacity since it also gives the
scattering rate of a photon by free electrons. Finally, $R$ denotes
the ``baryon-to-photon ratio,'' that is,
\begin{equation}
R \equiv \frac{3}{4} \frac{\rho_\BAR}{\rho_\gamma} .
\end{equation}
This factor in the baryon Euler equation ensures that the overall
momentum is conserved for the two fluids.

\subsubsection{The growth of Dark Matter perturbations}
\label{subsubsecdm}

Usually, Dark Matter is not coupled to any species (except through
gravitational interactions) so that its perturbations follow the
simple equation
\begin{equation}
\label{vcdmnor}
\dot v_\DM = - \Hconf v_\DM + k \Phi .
\end{equation}
It is well known that in the matter dominated epoch, the above
equation implies that Dark Matter density perturbations grow as the
scale factor: $\delta_\DM \propto a$.

In the radiation dominated epoch, it is also well known that the Dark
Matter density contrast [as soon as the fluctuation enters into the
Hubble radius (say $k > \Hconf = \eta^{-1}$)] grows
logarithmically~\cite{padm} as
\begin{equation}
\delta_{\DM} \sim \delta_k B(k\eta) ,
\label{deldmni}
\end{equation}
where
\begin{equation}
\label{eqB}
B (k\eta) \sim 1 - \alpha \ln (k\eta) + \beta \ln^2 (k\eta) ,
\end{equation}	
with $\alpha = v_k / \delta_k \sim 0$ and $\beta = \Phi_k / \delta_k
\sim 1$, where $\delta_k, v_k, \Phi_k$ are the amplitudes of the
corresponding quantities at $k \eta \ll 1$ (i.e., well before entering 
the Hubble radius; the actual values of $\alpha$ and $\beta$ actually 
depend on the initial conditions given, e.g., by inflation; see, for
example,~\cite{rl99}).  This yields approximately $B (k\eta) \sim 1 +
\ln^2 (k\eta)$.

It is particularly important to compute this quantity since the
growth~(\ref{deldmni}) will be suppressed when we consider the case
where Dark Matter is coupled to photons when it enters into the Hubble
radius.

\subsubsection{Free streaming and collisional damping}
\label{subsubsecdamp}

For standard cosmological perturbations, there are essentially two
damping phenomena: one (which mainly concerns neutrinos) is free
streaming, and the other one (which concerns photons) is collisional
damping.  Both of these are related to the presence of a small but
nonzero anisotropic stress in Eq.~(\ref{vphotex}). In order to compute
this anisotropic stress, one must remember that the usual Euler
equation is in fact part of the hierarchy of the Boltzmann equation in
which one expands the angular dependence of the temperature contrast
$\Theta$ in terms of multipoles, which obey the following hierarchy:
\begin{eqnarray}
\label{boltz}
\dot \Theta_\ell
 & = &   \frac{k}{2 \ell + 1} 
         [\ell \Theta_{\ell - 1} - (\ell + 1) \Theta_{\ell + 1}]
       + S_\ell - \dot \kappa \Theta_\ell .
\end{eqnarray}
This hierarchy involves source terms $S_\ell$, which include gravity,
polarization, etc.~\cite{pertgen1,pertgen2,pertgen3}. For $\ell =
1$, by comparing Eq.~(\ref{boltz}) to Eq.~(\ref{vphotex}), we easily
recover
\begin{eqnarray}
\Theta_0 & = & \delta_\gamma , \\
\Theta_1 & = & \frac{4}{3} v_\gamma , \\
\Theta_2 & = & \frac{1}{4} \pi_\gamma .
\end{eqnarray}
Photon free streaming occurs when both Thomson scattering and the
source terms are negligible. In this case, the hierarchy admits a
simple solution involving spherical Bessel functions:
\begin{eqnarray}
\Theta_\ell
 \propto j_\ell (k \eta)
 \propto \frac{\sin (k \eta + \phi)}{k \eta} , \quad k \eta \gg 1.
\end{eqnarray}
The interpretation of this behavior is that, since the mean free path
of the photons is very large, they simply flow away from the
overdense region toward the underdense regions. 

On the contrary, collisional damping appears when photons and baryons
are strongly coupled. For $\ell = 2$, Eq.~(\ref{boltz}) becomes
\begin{equation}
\dot \Theta_2
 =   \frac{k}{5} (2 \Theta_{1} - 3 \Theta_{3})
   + S_{2} - \dot \kappa \Theta_2 .
\end{equation}
For higher multipoles strong coupling implies 
that for $\ell > 2$, $\Theta_{\ell} \propto \dot \kappa^{- (\ell -
1)}$. For $\ell = 2$, if we can neglect $S_2$, we have
\begin{equation}
\Theta_2 = \frac{2}{5} \frac{k}{\dot \kappa} \Theta_{1} ,
\end{equation}
(If one takes polarization into account, then $S_2$ is of the same order
of magnitude as $\dot \kappa \Theta_2$, and the above two equations
are modified by small numerical factors.) Injecting this result into
Eq.~(\ref{vphotex}) and also taking into account the Boltzmann
equation for $\ell = 0$, one obtains a damped oscillator equation, in
which the term involving $\pi_\gamma$ acts as a damping term, which
takes into account both viscosity and heat conduction. This damping is
seen to be exponential and appears only when $k^2 \gg \dot \kappa /
\eta$. The interpretation of the above limit is that the damping occurs
only for scales smaller than the photon diffusion length. In
conclusion, the photon density perturbations follow
\begin{equation}
\delta_\gamma \sim \delta_k
\cos (k \eta / \sqrt{3})  e^{-\frac{2 k^2 \eta}{15 \dot \kappa}} .
\label{delgamel}
\end{equation}
(We have taken the limit $R \to 0$ here and neglected the influence of
polarization; see Refs.~\cite{hu1,hu2} for more detailed
calculations.)

Cosmological baryon density perturbations will essentially follow the
evolution of photon fluctuations as long as strong coupling is
effective. At sufficiently small scales, they will therefore be damped
before recombination. However, they will rapidly fall into the Dark
Matter potential well afterward, so that any damping in the baryon
fluid is rapidly ``forgotten'' after recombination. Of course, this
occurs only because there is an ``extra'' fluid (Dark Matter) which
was never coupled to photons and which can subsequently have some
gravitational interactions with baryons.

We shall now investigate how these conclusions are modified by the
presence of Interacting Dark Matter.

\subsection{Perturbation equations for IDM}
\label{subsecperteqicdm}

Assuming IDM is always nonrelativistic during the epochs of
interest, and interacts with photons only, we have the following
modified Euler equations for baryons, photons, and Dark Matter:
\begin{eqnarray}
\label{vbar}
\dot v_\BAR 
 & = &   k \Phi - \Hconf v_\BAR
       - R^{- 1} \dot \kappa (v_\BAR - v_\gamma) , \\
\label{vphot}
\dot v_\gamma
 & = &   k \Phi
       + \frac{1}{4} k \delta_\gamma - \frac{1}{6} k \pi_\gamma \nonumber \\
 &   & - \dot \kappa (v_\gamma - v_\BAR) 
       - \dot \mu (v_\gamma - v_\DM) , \\
\label{euler_DM}
\label{vcdm}
\dot v_\DM
 & = &   k \Phi - \Hconf v_\DM
       - S^{- 1} \dot \mu (v_\DM - v_\gamma) , 
\end{eqnarray}
where we have set
\begin{equation}
S \equiv \frac{3}{4} \frac{\rho_\DM}{\rho_\gamma}  
\end{equation}
(note that we also have $S \propto a$ and that $1 + R^{-1}, 1 + S^{-1}
\propto a^{-1}$ at early times and $\sim 1$ at late times), and where
$\dot \mu$ represents the interaction rate between photons and Dark
Matter. By analogy with Eq.~(\ref{kappa}), we write
\begin{equation}
\dot \mu \equiv a \sigma_\SGAMMADM n_\DM ,
\end{equation}
where $n_\DM = \rho_\DM / m_\DM$ is the Dark Matter number density and
$\sigma_\SGAMMADM$ is the photon--Dark-Matter cross section. For
simplicity, we shall assume that it is constant at low energy as for
the Thomson scattering case.  Since both Dark Matter and baryons are
supposed to be nonrelativistic, both $\dot \mu$ and $\dot \kappa$
behave as $a^{- 2}$ at high redshift. Their ratio is therefore
constant in this regime, and is characterized by the parameter
\begin{equation} 
u \equiv 
         \left[\frac{\sigma_\SGAMMADM}{\sigma_\THM} \right]
         \left[\frac{m_\DM}{100 \UUNIT{GeV}{}} \right]^{- 1} .
\end{equation}
This reads $\dot \mu = u \dot \kappa \Omega_\DM / 106 \Omega_\BAR$, or
$\dot \mu \sim \frac{1}{16} u \dot \kappa$ with our choice of
cosmological parameters.  We emphasize that the above parameter $u$
(measuring the relative size of $\dot \mu$ and $\dot \kappa$) is
defined before recombination. Note that after recombination $\dot
\kappa$ is strongly suppressed (by a factor $\sim 10^{-4}$,
see~\cite{peebles}) because of the drastic subsequent drop in the free
electron density, while we assume that $\dot \mu$ never suffers from
such a modification.

We shall now investigate the damping experienced by the Dark Matter
perturbations because of their coupling with photons.

\subsection{Dark-Matter--photon coupling}
\label{subsecdmgamma}

For many cases of interest, the differential opacity $\dot \mu$ is
large compared to the wavenumber $k$. In other words, the photon mean
free path is small compared to the scale of interest. This is true, in
particular, at high redshift, because $\dot \mu$ grows as $(1 + z)^2$.
In practice, one obtains a stiff system of equations, which physically
means that the relative velocity between the two species is small, so
that they can be considered as a single fluid. It is therefore more
convenient to consider the following two quantities:
\begin{eqnarray}
v_{\gamma\DM} & \equiv & \frac{v_\gamma + S v_\DM}{1 + S} , \\
\DD_{\gamma\DM} & \equiv & v_\gamma - v_\DM .
\end{eqnarray}
In term of these two variables, we have
\begin{eqnarray}
v_\gamma & = & v_{\gamma\DM} + \frac{S}{1 + S} \DD_{\gamma\DM} , \\
v_\DM & = & v_{\gamma\DM} - \frac{1}{1 + S} \DD_{\gamma\DM} .
\end{eqnarray}
Keeping in mind that $S \propto a$, so that $\dot S = \Hconf S$, we
then have
\begin{eqnarray}
\label{gvgb}
\dot v_{\gamma\DM}
 & = &   k \Phi
       + \frac{k}{1 + S} 
         \left(\frac{1}{4} \delta_\gamma - \frac{1}{6} \pi_\gamma \right)
       - \frac{S}{1 + S} \Hconf v_{\gamma\DM} 
\nonumber \\ & &
       - \frac{\dot \kappa}{1 + S} (v_\gamma - v_\BAR) , \\
\label{dvgb}
\dot \DD_{\gamma\DM}
 & = &   \frac{1}{4} k \delta_\gamma - \frac{1}{6} k \pi_\gamma 
       - \dot \kappa (v_\gamma - v_\BAR) 
\nonumber \\ & &
       - \frac{1 + S}{S} \dot \mu \DD_{\gamma\DM} .
\end{eqnarray}
In order to study the evolution of these quantities, we have to
consider various cases, depending on the relative values of $k$,
$\Hconf$, and $(1 + S^{- 1}) \dot \mu$. For a given wavelength, these
various case occur in the following chronological order:
\begin{itemize}

\item {\bf Case~1: Large wavelength limit}. This occurs when a given
  wavelength has not yet entered into the Hubble radius:
  \begin{equation} 
  k < \Hconf . 
  \end{equation} 
  In this case the cosmological perturbations do not experience any
  significant evolution, whatever the amplitude of the scattering
  is. Large wavelengths eventually enter into the Hubble radius after
  Dark Matter has decoupled from photons. In this case, we switch
  directly to case~4 below. Otherwise, we have case~2.

\item {\bf Case~2: Strong coupling regime}. This occurs when the
  scattering rate is higher than the expansion rate and the photon
  oscillation frequency: 
  \begin{equation} 
  \label{strongc}
  \Hconf < k < (1 + S^{-1}) \dot \mu .  
  \end{equation} 
  In this case,  Eq.~(\ref{dvgb}) reduces to 
  \begin{eqnarray}
  \DD_{\gamma\DM}
   & = & \frac{1}{(1 + S^{-1})} \frac{k}{\dot \mu}
  \nonumber \\ & & 
        \times \left(  \frac{1}{4} \delta_\gamma - \frac{1}{6} \pi_\gamma
                     - \frac{\dot \kappa}{k} (v_\gamma - v_\BAR) \right) .
 \end{eqnarray}
  The anisotropic stress can be neglected in the above equation,
  whereas in Eq.~(\ref{gvgb}) one has 
  \begin{equation}
  \label{pigamma} 
  \pi_\gamma = \frac{8}{5} \frac{k}{\dot \kappa + \dot \mu} v_\gamma . 
  \end{equation} 
  This ensures that the bulk velocities of the two fluids are almost
  identical, so that $v_{\gamma\DM}$ can be replaced by $v_\gamma$ in
  Eq.~(\ref{gvgb}). Microphysics plays a significant role in the
  evolution of the cosmological perturbations.  Thus, although small
  [see Eq.(\ref{pigamma})], the photon anisotropic stress cannot be
  neglected. In particular, this is the regime in which the photon
  fluctuations---with their Silk damping---are fully transferred to
  the Dark Matter.

\item {\bf Case~3: Weak coupling regime}. This represents the
  intermediate case, where \begin{equation} \Hconf < (1 + S^{-1}) \dot
  \mu < k . \end{equation} This regime always occurs between the
  strong coupling regime and decoupling since the interaction rate
  must drop below $k$ before reaching $\Hconf$. This regime is really
  effective when it lasts several expansion times, and gives rise to
  new, unexpected effects.
  
  This regime, indeed, is new: for baryons coupled to photons, it is
  not relevant. Thomson scattering gives rise to the same succession
  of events: $k < \Hconf$, $\Hconf < k < (1 + R^{- 1}) \dot
  \kappa$. However, in this case, for the relevant wavelengths,
  recombination occurs in the strong coupling regime. Thus, within a
  negligible fraction of time, one switches to $(1 + R^{- 1}) \dot
  \kappa < \Hconf < k$, therefore skipping the weak coupling
  regime. Moreover, the damping due to the weak coupling regime is due
  to an averaging of the photon fluctuations, which are transmitted to
  the Dark Matter, over many oscillations. This requires in the baryon
  case $k \delta \tau \gg 1$, where $\delta \tau $ is the thickness of
  the last scattering surface. In this regime, many other sources of
  damping are present, compared to which weak coupling effects are
  negligible. The weak coupling regime for baryons does not occur for
  waves of horizon size such as the ones considered for Dark Matter.

\item {\bf Case~4: No coupling}. This occurs when the scattering rate
  is negligible with respect to the expansion rate: \begin{equation}
  \label{nocoupling} (1 + S^{-1}) \dot \mu < \Hconf < k .
  \end{equation} In this case, it is safe to neglect all the terms
  involving \DMGAMMA{} scattering in the above
  equations~(\ref{vphot},\ref{vcdm}) and the two fluids evolve
  independently from each other.

\end{itemize}

Clearly, the interesting cases to be discussed are cases~2 and~3.

\subsection{Dark-Matter--photon decoupling before recombination}
\label{subsecdamp2}

The (more interesting) case we will focus on in detail is when Dark
Matter is coupled to photons and the photon interactions are dominated
by the Thomson scattering process (e.g., when the photon--Dark-Matter
interactions decouple before recombination). We will also assume that
we are before the radiation to matter transition.  In this case, the
Dark Matter perturbations will be {\em driven} by the photon
perturbations, which follow their usual behavior.  As we suppose Dark
Matter decouples from photons before recombination, we have $u < 2$.
This implies that the photon coupling is stronger with baryons than
with Dark Matter.  This condition is valid for any realistic nonzero
Dark-Matter--photon interactions.  Depending on the amplitude of the
photon--Dark-Matter coupling (and, hence, on the epoch where they
decouple) several different effects are to be expected.  We are now
going to discuss cases 2 and 3 defined above (strong and weak
coupling).

For simplicity, we have considered only the case of a radiation
dominated Universe, with $R, S \ll 1$. This will be sufficient to
explain the important results: the only cases where this does not hold
are not relevant, because realistic models predict that Dark Matter
must decouple from the photons before equality~\cite{bfs2001}.  A more
complete classification may be found in Ref.~\cite{bfs2002}.  The
extension to the matter dominated case, if desired, is straightforward
anyway. Our numerical results are of course given without such a
restriction.

\subsubsection{Strong coupling regime ($\Hconf < k < S^{-1} \dot \mu$)}
\label{subsubrefstrong}

In this case, we have $\delta_\DM \sim \delta_\gamma$, with the photon
fluctuations given by their usual expression~(\ref{delgamel}).

\paragraph{Late decoupling.}

\begin{figure}
  \centerline{ \psfig{file=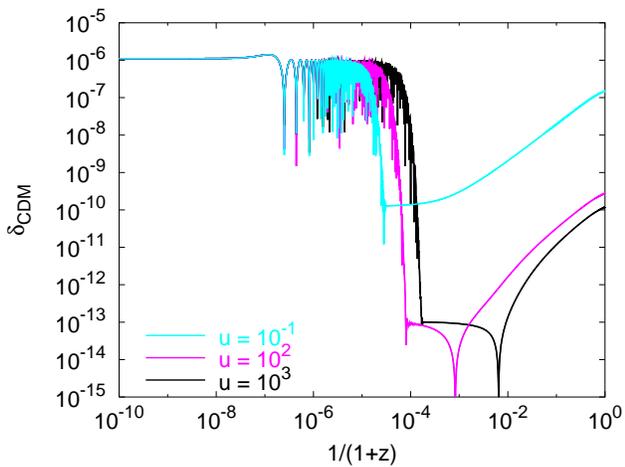,angle=270,width=3.5in}}
  \caption{Evolution of Dark Matter density perturbation as a function
  of the redshift for various ({\em large}) interaction rates with
  photons, all leading to collisional damping. In this plot, as in the
  others, we consider a Dark Matter model with a cosmological
  constant, with $h = 0.65$ (i.e., $H_0 = 100 h \UUNIT{km}{}
  \UUNIT{s}{-1} \UUNIT{Mpc}{-1}$), $\Omega_{\Lambda} = 0.7$,
  $\Omega_\MAT = 0.3$, $\Omega_\BAR h^2 = 0.019$, and a scalar
  perturbation index $n_\SCAL = 1$. We have considered the mode $k =
  40 \UUNIT{Mpc}{- 1}$. At $z > 10^7$, the mode is outside the Hubble
  radius and the perturbation is frozen. The perturbation first
  experiences undamped oscillations (strong coupling regime) and then
  exponentially damped oscillations, corresponding to the collisional
  damping regime. The \GAMMADM{} cross sections are parameterized by
  the quantity $u \equiv [\sigma_\SGAMMADM / \sigma_\THM ] [m_\DM /
  100 \UUNIT{GeV}{} ]^{- 1}$.}
\label{fig_cd}
\end{figure}

Collisional damping in the Dark Matter fluid occurs when the photons
are subject to damping, which is due---as usual---to the presence of
the baryons, when $k^2 \gg \dot \kappa / \eta$. This as previously
implies many oscillations after the mode $k$ enters into the Hubble
radius before damping is to occur (Fig.~\ref{fig_cd}).  Of course, as
one can see on Fig.~\ref{fig_cd}, the damping phenomenon starts later,
but lasts longer and is more important as the \GAMMADM{} cross section
increases. Note also (this will be important later) that the epoch at
which the collisional damping stops is difficult to compute. In the
case considered here, it occurs when $\Phi$ in
Eqs.~(\ref{vbar}--\ref{vcdm}), although quite small, is no longer
negligible compared to the strongly damped density contrasts. In other
words, the damping stops when gravity becomes the dominant term in the
Euler equation.

\paragraph{Early decoupling.}
\label{subsubsecsmall}

\begin{figure}
  \centerline{ \psfig{file=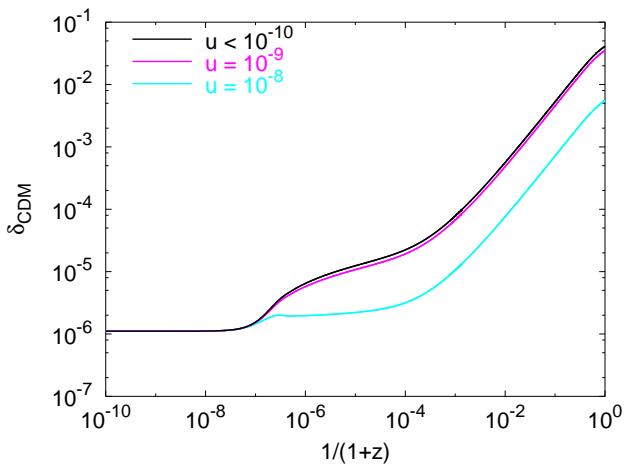,angle=270,width=3.5in}}
  \caption{Evolution of CDM density perturbation as a function of the
  redshift for various (small) interaction rates with photons. As
  expected, for sufficiently small cross sections, almost no effect is
  noticeable. For slightly larger cross sections, the coupling between
  Dark Matter and photons stops just after the mode has entered into
  the Hubble radius, preventing the perturbation from experiencing the
  small growth in the radiation dominated era.}
\label{fig_nc}
\end{figure}

As opposed to the above case, for much smaller cross section, there
still can be coupling between Dark Matter and photons, but without
collisional damping at the scales of interest. This occurs for
cross sections sufficiently small so that the decoupling occurs soon
after the mode enters into the Hubble radius, in which case only the
small logarithmic growth~(\ref{deldmni}) of Dark Matter perturbations
during the radiation dominated epoch can be suppressed. Some example
of this are shown on Fig.~\ref{fig_nc}. Note that the suppression of
the logarithmic growth during the radiation era is enough to reduce
the power spectrum of the Dark Matter fluctuations by one order of
magnitude, which is already a large effect.

\subsubsection{A new damping regime: The weak coupling regime
($\Hconf < S^{-1} \dot \mu < k$)}
\label{subsubsecnew}

When one considers cross sections intermediate to those considered in
the above paragraphs, a new phenomenon can occur. For this, several
conditions must be satisfied.
\begin{enumerate}

\item The coupling between Dark Matter and photons is sufficiently
  small so that we have $v_\DM \not\sim v_\gamma$.

\item The coupling between Dark Matter and photons is sufficiently
  large so that the velocity of Dark Matter perturbation is ``driven''
  by that in the photon perturbations: this means that they experience
  oscillation at the same frequency and therefore $\dot v_\DM \sim (k
  / \sqrt{3}) v_\DM$.

\item Gravity must be negligible in the Dark Matter Euler equation. 

\end{enumerate}
It is easy to see that this can occur in the above defined {\em weak
coupling regime}. Then, Eq.~(\ref{vcdm}) reduces to
\begin{equation}
\dot v_\DM \simeq S^{-1} \dot \mu v_\gamma \propto k v_\DM .
\label{wceq}
\end{equation}
This implies
\begin{equation}
\delta_\DM \sim 
 \frac{S^{-1}\dot\mu}{\Hconf} 
 \frac{\sin (k\eta/{\sqrt{3}})}{k\eta/{\sqrt{3}}} 
  e^{-\frac{2 k^2 \eta}{15 \dot \mu}} , \quad k \eta \gg 1 .
\label{delwc}
\end{equation}
Now, in the radiation dominated era, as long as there is no
collisional damping between photons and baryons, $v_\gamma$
experiences undamped oscillations. This means that {\em the Dark
Matter fluctuations are damped as $S^{-1} \dot \mu$ as a function of
time}, that is, as $a^{-3}$.

Except for a time-dependent normalization factor (which reduces to
unity at the \DMGAMMA{} decoupling), this form is very similar to the
damping due to the free streaming of a relativistic fluid.  Obviously,
in the present case, the fluids are coupled and far from free
streaming.  Both the photon and Dark Matter mean free paths are still
small; therefore we are well in the collisional regime. However, the
coupling rate between Dark Matter and photons $S^{- 1} \dot \mu$ is
much smaller than the photon oscillation frequency $k / \sqrt{3}$.
The slow reaction of the Dark Matter to the photon oscillations then
mixes modes with different phases, as does the free streaming process
which collects particles of different origin.
\begin{figure}
  \centerline{ \psfig{file=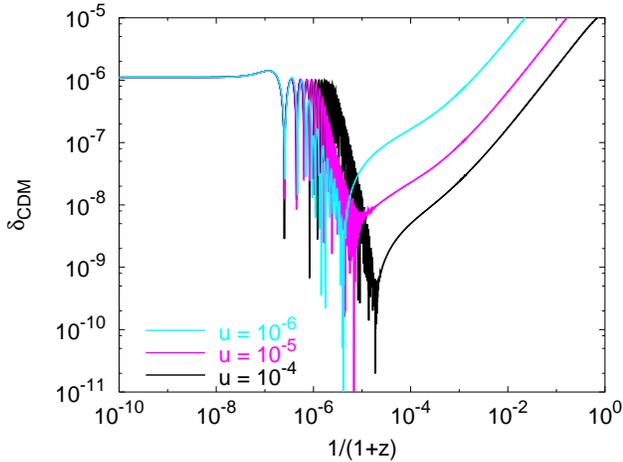,angle=270,width=3.5in}}
  \caption{Evolution of CDM density perturbation as a function of the
  redshift for various (intermediate) interaction rates with
  photons. As explained in the text, the Dark Matter perturbations
  experience a power law decay due to their weak coupling to
  photons. This is the main effect we can expect to have at small
  scales for acceptable cross sections.}
\label{fig_lc}
\end{figure}
Of course, should $v_\gamma$ be damped or affected by another fluid
for one reason or another, then the CDM fluctuations would also feel
it. For example, collisional damping in the photon--baryon fluid may 
already be at work in the weak coupling regime, or may also be effective
after weak coupling occurs and thus less apparent.  This is in fact
what we can barely see on Fig.~\ref{fig_lc} for the highest
cross section where the damping obviously increases soon before
decoupling.

To be fully developed, the weak coupling regime requires $k \gg
\Hconf$.  For $k \ge \Hconf$, which will be a case of importance
below, it is not well separated from the strong coupling regime. Due
to the rapid variations in time of $S^{-1} \dot \mu$ compared to
$\cos (k \eta / \sqrt{3})$ near $k \eta \sim 1$, Eq.~(\ref{delsc})
still holds during the first few oscillations.  The Dark Matter
fluctuations are thus given by Eq.~(\ref{delsc}) for $k \sim \Hconf$,
that is $k \eta \sim 1$, and go over to Eq.~(\ref{delwc}) for $k \eta
\gg 1$.

\subsection{Dark-Matter--photon decoupling after recombination}
\label{subsecdamp1}

In this nonstandard scheme, the photon--Dark-Matter interactions
decouple after the recombination epoch. This is therefore the case
where the usual coupling between photons and baryons is vanishingly
small after recombination as oppose to the coupling between photons
and Dark Matter. With our choice of cosmological parameters, this case
occurs for $u > 2$.  This unrealistic example is given here for
pedagogical purpose only. Clearly, collisional damping in this case is
larger than the Silk damping, implying that this scenario is of no
cosmological relevance.

As Dark Matter and photons are more tightly coupled than photons and
baryons, the damping of the fluctuations is there due to the
interaction of the photons with the Dark Matter, not with the
baryons. In this case the expression~(\ref{pigamma}) for $\pi_\gamma$
becomes
\begin{equation}
\label{pigamma2}
\pi_\gamma
 \sim \frac{k}{\dot \mu} \frac{8}{5} v_\gamma .
\end{equation}
Hence, after recombination (and also before recombination in case
$\dot \mu > \dot \kappa$, that is $u > 16$), we have
\begin{equation}
\delta_\DM \sim \delta_\gamma \sim  \delta_k
\cos ({k\eta} / {\sqrt{3}})  e^{-\frac{2 k^2 \eta}{15 \dot \mu}} .
\label{delsc}
\end{equation}
Note that even in this case, the damping starts only after many
oscillations as one must require $k^2 \gg \dot \mu / \eta$, which may
not yet hold when the mode enters into the Hubble
radius. Figure~\ref{fig_cd} shows some examples of Dark Matter and
photons experiencing collisional damping due to \GAMMADM{}
interactions.

\subsection{Dark Matter damping factor}
\label{subsecdamp3}

We may now evaluate the total damping at the decoupling of the Dark
Matter with the photons. Provided the fluctuations are not too much
damped (this is indeed the only relevant case: it is of no interest to
evaluate very accurately fluctuations which are negligible), it is
given by Eqs~(\ref{delgamel}) or~(\ref{delwc}) to which the
exponential collisional damping is added. The latter is to be taken at
the time where the Dark Matter decouples from the photons, namely,
$\eta = \eta_\DEC$, the solution of $S^{-1} \dot \mu \sim \Hconf$:
\begin{eqnarray}
\label{Tdm1}
T_\IDM & \sim &
 \frac{\cos ({k \eta_\DEC}/{\sqrt{3}})}{B(k \eta_\DEC)} 
 e^{-\frac{2 k^2 \eta_\DEC}{15 \dot \kappa}} 
,\quad
k \eta_\DEC  \sim  1 , \\
T_\IDM & \sim &
 \frac{\sin (k \eta_\DEC / {\sqrt{3}})}
      {B (k \eta_\DEC) k \eta_\DEC / {\sqrt{3}}}
 e^{-\frac{2 k^2 \eta_\DEC}{15 \dot \kappa}}
,\quad
k\eta_\DEC  \gg  1 .
\end{eqnarray}
The preexponential oscillating factor is the reduction due to the
strong and weak coupling to the photons, and $B (k\eta_\DEC)$ is the
(logarithmic) reduction due to the impeded growth of the Dark Matter
fluctuations compared to the noninteracting case.  These factors come
into play as soon as $k \eta_\DEC$ is larger than unity, that is, for
the modes which just enter the Hubble radius at the \DMGAMMA{}
decoupling.  The exponential factor is the damping due to viscous
effects.  It enters into play for modes which are of the size of the
length traveled by the collisional photons at the time of the
\DMGAMMA{} decoupling. Due to the strong \GAMMAE{} interaction,
this length is in the present case much smaller than the Hubble
radius.

\subsection{CMB anisotropies}
\label{subsecCMB}

\begin{figure}
  \centerline{ \psfig{file=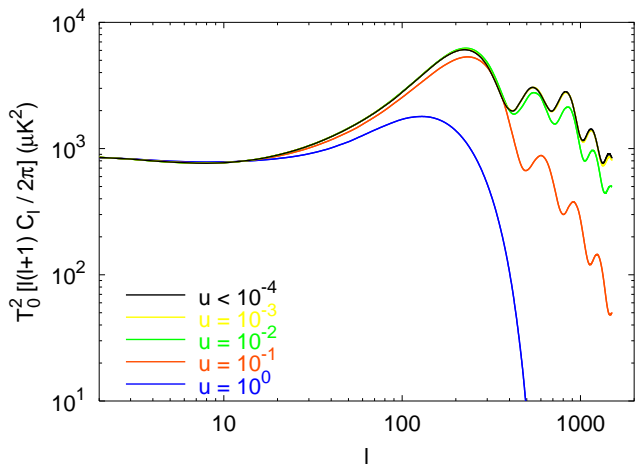,angle=270,width=3.5in}}
  \caption{Influence of Interacting Dark Matter on the CMB anisotropy
  spectrum as a function of the Dark-Matter--photon cross section.
  The most spectacular effect which occurs at sufficiently large
  $\sigma_\SGAMMADM$ is the apparent damping due to the large width of
  the last scattering surface, with some additional collisional
  damping due to photon--Dark-Matter interactions, and a slight shift
  of the Doppler peaks due to the decreases of the sound speed in the
  Dark-Matter--baryon--photon ``plasma''.  }
\label{fig_b4}
\end{figure}

When computing the CMB anisotropies, one must take into account the
modification to the photon Boltzmann hierarchy induced by the Dark
Matter interactions. This is due to the fact that both free electrons
and Dark Matter particles are responsible for photon scattering.  For
the scalar part of the multipoles $\Theta_\ell$ of the distribution
function, one now has 
\begin{eqnarray}
\dot \Theta_\ell
 =   \frac{k}{2 \ell + 1} 
     [\ell \Theta_{\ell - 1} - (\ell + 1) \Theta_{\ell + 1}]
   + S_\ell - (\dot \kappa + \dot \mu) \Theta_\ell ,
\end{eqnarray}
where $S_\ell$ is the usual source term which also involves an extra
term involving the Dark Matter velocity, as photons can scatter on
both baryons and Dark Matter.  Obviously, the differential opacity is
now $\dot \kappa + \dot \mu$, which can in some cases be different
from the usual term $\dot \kappa$. In this case, recombination and the
subsequent drop in the free electron density do not necessarily imply
photon decoupling because of their interactions with Dark
Matter. Therefore, large interactions between Dark Matter and photons
can significantly delay the epoch of photon last scattering.  The net
effect will be to enlarge the width of the last scattering surface,
and hence to increase the damping of the observed CMB anisotropies.
Of course, the effect is qualitatively similar to late reionization.
The presence of several Doppler peakw as detected by the most recent
experiments~\cite{CMBnew1,CMBnew2,CMBnew3} therefore puts a firm upper
limit on the Dark-Matter--photon cross section: its contribution to
the differential opacity must be small at $z \sim 1000$. Some examples
are represented on Fig.~\ref{fig_b4}.  We therefore have by eye the
constraint
\begin{equation}
\frac{\sigma_\SGAMMADM}{m_\DM}
 \lesssim 10^{-3} \frac{\sigma_\THM}{100 \UUNIT{GeV}{}} .
\end{equation}
Actually, this is roughly speaking the constraint that one must have
in order not to significantly increase the opacity {\em just after}
recombination.

In addition, some other effects are in principle observable on the CMB
anisotropy spectrum.
\begin{itemize}
  
\item First, one expects that there will be collisional damping on
  small scales, which will also produce an exponential cutoff in the
  spectrum (this will be discussed later). For realistic cross
  sections, which could damp perturbations below $100 \UUNIT{kpc}{}
  \sim 10^{-3} 100 \UUNIT{Mpc}{}$, this occurs roughly at angular
  scales $\sim 10^3$ smaller that the first Doppler peak, i.e., around
  $\ell \sim 10^5$ ! This is of course far too small to be observable
  even in the far future.  For higher cross sections, the effect can
  be similar to that of large width of the last scattering surface.
  Distinguishing between the two is not easy, but it happens that the
  latter is dominant in our model.
  
\item Second, the sound speed is modified by the presence of Dark
  Matter. The sound speed is now 
  \begin{equation} c_\SOUND =
  \frac{1}{\sqrt{3}} \frac{1}{(1 + R + S)^{\frac{1}{2}}} ,
  \end{equation} 
  instead of $[3 (1 + R)]^{-\frac{1}{2}}$ in the case of strong
  coupling.  This means that if Dark Matter is still coupled to
  photons at last scattering, the acoustic oscillation will have a
  lower frequency and the Doppler peak structure will be shifted to
  higher multipoles. This is what we can see on Fig.~\ref{fig_b4} for
  $\sigma_\SGAMMADM = \UEE{-1}, \UEE{-2} \sigma_\THM$. For lower cross
  sections, Dark Matter has already decoupled at $z \sim 1000$, and
  for higher cross sections, the exponential damping also
  significantly shifts the peak positions in the other direction. It
  seems that one cannot easily shift the peak position in that way
  without also modifying the photon last scattering history. This is
  due to the fact that there is {\it a priori} no reason for a violent
  drop in $\dot \mu$ around $z \sim 1000$ as is the case for $\dot
  \kappa$ (see also~\cite{dub}).

\end{itemize}
It happens that these effects are unobservable in realistic models
given the much stronger constraints that arise from the matter power
spectrum.

\subsection{Matter power spectrum}
\label{subsecPS}

\begin{figure}
  \centerline{ \psfig{file=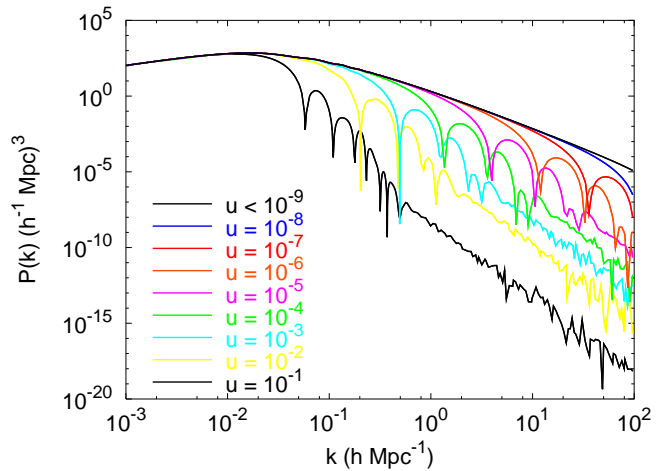,angle=270,width=3.5in}}
  \caption{Influence of Interacting Dark Matter on the matter power
  spectrum. As explained in the text, the deviation from a CDM power
  spectrum exhibits several regimes, the strong coupling regime, the
  collisional damping regime, and the neutrino regime. The cosmological
  parameters here are the same as for Fig.~\ref{fig_b4}.  The wiggles
  at very small scales for large cross section are due to some
  unimportant numerical accuracy problems.}
\label{fig_b5}
\end{figure}

When studying the influence of IDM on the matter power spectrum, one
can expect to observe four different regimes.
\begin{itemize}
  
\item For small $k$ (large wavelengths), the perturbations enter the
  Hubble radius rather late, when $k = \Hconf \gg S^{- 1} \dot \mu$.
  This occurs after the $\dot \mu$ terms of
  Eqs.~(\ref{vphot},\ref{vcdm}) have already become negligible. So the
  mode enters into the Hubble radius decoupled.  For these modes,
  there is no difference from the usual case where there is no
  coupling between Dark Matter and photons.
  
\item For larger $k$ (smaller wavelengths), the mode enters
  the Hubble radius when $k = \Hconf \ll S^{- 1} \dot \mu$, when the
  Dark Matter is coupled to the photons.  This results in a reduction
  in the Dark Matter fluctuation amplitude compared to the
  noninteracting case as soon as the mode is within the Hubble radius.
  If the interaction of the Dark Matter with the photons is not too
  strong, the \DMGAMMA{} decoupling occurs before collisional damping
  is sizable.  This corresponds to the {\em weak coupling regime}.
  The spectrum then shows a characteristic behavior due to this weak
  coupling, namely, a series of damped oscillations of slope
  $k^{n_\SCAL - 6}$.
  
\item For even larger $k$ (smaller scales), the perturbation enters
  the Hubble radius even earlier, still when $k = \Hconf \ll S^{- 1}
  \dot \mu$, but also when $k$ is sufficiently large so that at a
  later time $k^2 / \dot \kappa \gg \Hconf$ before the
  Dark-Matter--photon decoupling.  Then, the perturbations experience
  collisional damping: Dark Matter, photon and/or baryon perturbations
  are exponentially damped. This translates into an exponential cutoff
  in the matter power spectrum.  This occurs for larger \GAMMADM{}
  cross sections or smaller scales than the previous regime.  We call
  this the {\em collisional regime}.  So there is a range of
  interaction rates and scales where only the previous regime provides
  the damping.
  
\item Finally, at very small scales, a new behavior appears: when the
  Dark-Matter--baryon--photon perturbations are enormously damped, the
  only significant perturbations that survive are those of
  (relativistic) neutrinos (this could be called the {\em neutrino
  regime}).  These of course have also experienced significant damping
  because of free streaming. The latter, however, is much less
  effective than the collisional damping. The neutrino fluctuations
  can therefore eventually dominate.  Through gravity, they regenerate
  Dark Matter fluctuations which, although quite small, are much
  larger than they would be under the sole action of collisional
  damping.  For relativistic species, the damping of the density
  perturbations varies only as $(k \eta)^{-1}$ ($\eta$ being the
  conformal time), which explains the $k^{n_\SCAL - 6}$ slope at high
  $k$ on Fig~\ref{fig_b5}.  This undoubtedly occurs when the
  amplitudes are extremely small, and anyway yields a contribution
  which for most applications is negligible compared to the one at
  larger scales.  Note that such a behavior can also be seen in a
  model when all the (non interacting) Dark Matter is made of one or
  two massive neutrino species.

\end{itemize}
These four regimes can easily be seen on Fig.~\ref{fig_b5} for various
\GAMMADM{} cross sections.

\section{Reduction of small scale power}
\label{steen_part}

The reduction of power on small scales for various Dark Matter
candidates can be described by a transfer function $T_X$ defined such
that
\begin{equation}
P_X = T^2_X \cdot P_\CDM  ,
\end{equation}
where $P_\CDM$ is the corresponding Cold Dark Matter power.  For non
interacting WDM particles, this damping due to free streaming is
traditionally set, by convention, at masses $M < 10^{12} M_\odot$, and
is described in Ref.~\cite{bardeen} by an exponential cutoff $T_\WDM =
\exp [-k R_f / 2 - (k R_f)^2 / 2]$, where the comoving free streaming
scale $R_f$ is given in terms of the Dark Matter
mass~\cite{sommerlarsen}, $R_f = 0.2 (\Omega_\WDM h^2)^{1/3} \left(
m_\WDM / 1 \UUNIT{keV}{} \right)^{-4/3} \UUNIT{Mpc}{}$.  A somewhat
more accurate result is found from a Boltzmann code calculation,
giving~\cite{bode}
\begin{equation}
T_\WDM = [1 + (\alpha k) ^{2 \nu}]^{-5 / \nu} ,
\label{fitmass}
\end{equation}
where 
\begin{eqnarray}
\alpha & = & 0.048 \UUNIT{Mpc}{} (\UUNIT{keV}{}/m_\DM)^{1.15}  
\nonumber \\
       & & \times (\Omega_\DM /0.4)^{0.15} (h/0.65)^{1.3}
       (1.5/g_\DM)^{0.29} ,
\end{eqnarray}
with $\nu = 1.2$, and $g_\DM = 1.5$ for a neutrinolike Dark Matter
candidate. For specific sterile neutrino WDM candidates this form is
somewhat changed~\cite{Hansen:2001zv}.

{}From Fig.~\ref{fig_b5}, it is clear that IDM can provide an initial
reduction of small scale power similar to what WDM gives.  A good fit
to the transfer function
\begin{equation}
P_\IDM = T^2_\IDM (k) \cdot P_\CDM ,
\end{equation}
is (at least near $u = [\sigma_\DM / \sigma_\THM] [m_\DM / 100
\UUNIT{GeV}{}]^{-1} \sim \UEE{-6}$)
\begin{equation}
T_\IDM = [1 + (\alpha k)^{2 \nu}]^{-5 / \nu} ,
\label{fitsigma}
\end{equation}
where again $\nu = 1.2$ and
\begin{equation}
\alpha = 0.073 \UUNIT{Mpc}{} \left( u/10^{-6} \right)^{0.48} .
\end{equation}
On smaller scales this expression naturally breaks down because of the
presence of oscillations in the spectrum, but we are mostly interested
in the scale at which IDM begins to produce significant deviations
from the standard CDM case. It is justified since the second maximum
is already down by at least an order of magnitude.\footnote{This
procedure amounts to fitting the {\em first downward oscillation},
i.e., the moment $k \eta_\DEC \sim 1$ in Eq.~(\ref{Tdm1}).  Indeed, we
see that $\alpha$ is nearly proportional to $\eta_\DEC$.  Our
cosmological parameters imply $\eta_\DEC = 0.35 \UUNIT{Mpc}{} \left(u
/ \UEE{-6} \right)^{0.5}$, that is, $\alpha k \sim k \eta_\DEC /
5$. Undoubtedly, the fitting form~(\ref{fitsigma}) just reproduces the
fall-off of $\cos (k \eta_\DEC / \sqrt{3})$ damped by the factor $B (k
\eta_\DEC)$, Eqs.~(\ref{eqB},\ref{Tdm1}).  In the case of free streaming,
for a damped sinusoidal function, quite a similar fall-off is present,
but for totally different reasons. This nevertheless explains the
similarity of the fitting forms.}

A comparison of Eqs.~(\ref{fitmass}) and~(\ref{fitsigma}) reveals
that a heavy particle (e.g., $m = 100 \UUNIT{GeV}{}$) with scattering
cross section with photons of approximately $\sigma_\SGAMMADM =
10^{-6} \sigma_\THM$, can provide the same reduction of small scale
power as a conventional WDM particle with mass $m = 1
\UUNIT{keV}{}$. We thus see explicitly how IDM can disguise itself as
WDM.

On the other hand, one immediately gets constraints on the allowed
scattering cross section for the following reason. In order to
reproduce the observed properties of the Lyman-$\alpha$ forest in
quasar spectra one gets a bound on the free streaming
scale~\cite{narayanan}, corresponding to a WDM mass of approximately
$0.75 \UUNIT{keV}{}$.  Furthermore, by extending the Press-Schechter
formalism to include WDM (see, e.g., \cite{ss1988}, where the problems
this raises are discussed), one can study galaxy formation with
varying WDM mass. Combined with the existence of a supermassive black
hole at $z = 5.8$ one finds~\cite{barkana} a lower bound on the WDM
mass of approximately $0.75 \UUNIT{keV}{}$. These results apply
equally well to IDM, because of the damping of small scale power, and
the results of Refs.~\cite{narayanan,barkana} translate into the bound
\begin{equation}
\frac{\sigma_\SGAMMADM}{m_\DM}
 \lesssim 10^{-6} \frac{\sigma_\THM}{100 \UUNIT{GeV}{}} \simeq
 \EE{6}{-33} \UUNIT{cm}{2} \UUNIT{GeV}{-1} .
\end{equation}
This bound is stronger than the one, $u < \EE{5}{-4}$~\cite{bfs2001},
obtained by considering collisional damping alone. The reason is the
reduction in the amplitude of the Dark Matter fluctuations, coupled to
the photons before collisional damping sets in.  As already mentioned
this bound is also stronger than the constraint $u < 10^{-3}$ arising
from CMB anisotropies.

This bound is obtained under the conservative assumption that the
abundance of $10^{12} M_\odot$ galaxies is correctly given by CDM, and
requires no knowledge of the number of smaller objects. Should we take
the more incisive point of view that CDM yields the observed abundance
of $10^{9} M_\odot$ objects, then the above bound is lowered by
somewhat more than an order of magnitude.

\section{Conclusion}

We have considered the effect of Dark-Matter--photon interactions on
the evolution of primordial Dark Matter fluctuations.  Rather than
growing when they enter the horizon, the fluctuations stay of constant
amplitude, as do the photon fluctuations. This impeded growth appears
as a damping compared to the original amplitude of the
fluctuations. Moreover, fluctuations on scales much below the size of
the horizon are seen to couple to the photons at a rate much lower
than the rate at which they oscillate. This is an additional, new,
damping process we have called weak coupling. As a result,
horizon-size Dark Matter fluctuations are seen to be damped if coupled
to photons.  The usual (exponential) damping sets in for much smaller
scales.  We hence have obtained a new constraint on the allowed cross
sections.  This bound~(\ref{wcbound}) is two orders of magnitude
stronger than the necessary condition obtained by considering the
exponential damping of Dark Matter fluctuations induced by the photon
interactions~\cite{bfs2001}.  This new bound reads
\begin{equation}
\label{wcbound}
\sigma_\SGAMMADM / m_\DM \lesssim 10^{-32} \UUNIT{cm}{2} \UUNIT{GeV}{-1} .
\end{equation} 
The maximum allowed value reduces the matter power spectrum in a way
corresponding to a conventional Warm Dark Matter particle with a mass
of about $1 \UUNIT{keV}{}$ but leaves the cosmic microwave background
anisotropies undisturbed.  Using recent bounds on the scale of
reduction of the matter power spectrum thus allows us to put new
bounds on the allowed photon--Dark-Matter cross section.

This corresponds to a universe which is well transparent to photons:
the free mean path of a photon due to the interactions with Dark
Matter in a halo core of mass density $0.02 M_\odot \UUNIT{pc}{- 3}$
is of order $\EE{6}{4} \UUNIT{Gpc}{}$, and the optical thickness
toward the (usual) last scattering surface is below $10^{- 5}$.  The
value~(\ref{wcbound}) nevertheless remains quite large compared to the
theoretical estimates usually encountered for weakly interacting
particles, although there are no compelling reasons to exclude
it. Anyway, this leaves open new possibilities as far as the nature of
Dark Matter is concerned.

The lower bound~(\ref{wcbound}) implies that Dark Matter decouples
from the photons before the collisional Silk damping is at work,
leaving an oscillating, power-law, damped matter power spectrum.  This
new damping regime bears some similarities with the free streaming
case, although here the Dark Matter and photon fluids are undoubtedly
coupled. Such Dark Matter particles therefore appear to be good Warm
Dark Matter candidates, with features in the matter power spectrum
different from the conventional WDM at very small scales.

\acknowledgments

The authors wish to thank Jim Bartlett, Julien Devriendt, Pierre
Fayet, G\'erard Mennessier, Gilbert Moultaka, and James Taylor for
enlightening discussions, and Institut d'Astrophysique de Paris where
part of this work was completed.  C.B. is supported by a PPARC
Fellowship. A.R. was funded by EC Research Training Network CMBNET
(contract number HPRN-CT-2000-00124) at the beginning of this work.
S.H.H. is supported by a Marie Curie Fellowship of the European
Community under the contract HPMF-CT-2000-00607.  This work was
initiated thanks to the French Groupement de Recherches sur la
Supersym\'etrie.

\end{document}